\documentclass[%
pre,
aps,
aip,%
amsmath,amssymb,floatfix,
reprint,%
twocolumn,%
jcp,%
longbibliography,
reprint,%
]{revtex4-1} 

\usepackage{amsmath,amsthm,latexsym,amssymb,amsfonts,epsfig}

\usepackage[english]{babel}
\usepackage[utf8]{inputenc}			
\usepackage{graphicx, texdraw}  
\usepackage{color}
\usepackage{bm}

\usepackage{type1ec}

\usepackage{mathtools}
\usepackage{mathrsfs} 

\usepackage[left= 1.7cm, right = 1.7cm, top= 2.3cm, bottom = 2.3cm]{geometry}  

\usepackage{enumerate}

\usepackage[colorlinks=true,citecolor=OliveGreen, linkcolor=blue]{hyperref}

\usepackage{csquotes} 

\usepackage{color}
\usepackage[svgnames,dvipsnames]{xcolor}

\usepackage{amsmath}
\usepackage{amsfonts}
\usepackage{amssymb}

\newcommand{\vect}[1]{\bm{#1}}

\newcommand{\bv}{\vect{v}}

\newcommand{\boldNabla}{\boldsymbol{\nabla}}

\newcommand{\kS}{\kappa_\mathrm{S}}
\newcommand{\kA}{\kappa_\mathrm{A}}
\newcommand{\kB}{\kappa_\mathrm{B}}

\newcommand{\kBolt}{k_\mathrm{B}}

\newcommand{\betaB}{\beta_\mathrm{B}}

\newcommand{\G}{\mathcal{G}}
\newcommand{\R}{\vect{r}}
\newcommand{\Intd}{\mathrm{d }}
\newcommand{\PP}{\mathcal{P}}

\newcommand{\F}{\vect{F}}

\newcommand{\vt}{\tilde{v}}

\newcommand{\bigO}{\mathcal{O}}

\newcommand{\Faxen}{Fax\'{e}n}

\newcommand{\ETH}{\epsilon_{\mathrm{th}}}

\newcommand{\eZ}{\vect{e}_{z}}
\newcommand{\eX}{\vect{e}_{x}}
\newcommand{\eY}{\vect{e}_{y}}

\newcommand{\gOne}{\vect{g}_1}
\newcommand{\gTwo}{\vect{g}_2}

\newcommand{\JS}{J_{\mathrm{S}}}

\newcommand{\Gmatr}{\boldsymbol{\mathcal{G}}}

\newcommand{\EB}{E_\mathrm{B}}

\newcommand{\Rbig}{\vect{R}}

\newcommand{\bNabla}{\boldsymbol{\nabla}}

\newcommand{\mi}{\bm{\mu}}

\newcommand{\ie}{{\it i.e.~}}

\renewcommand{\cite}{\citep}

\usepackage{setspace}
\usepackage{multirow}

\usepackage{epstopdf}

\makeatletter
\def\l@subsubsection#1#2{} 

\def\l@@sections#1#2#3#4{%
 \begingroup
  \everypar{}%
  \set@tocdim@pagenum\@tempboxa{#4}%
  \def\testA{subsection}\def\testB{#2}\ifx\testA\testB \set@tocdim@pagenum\@tempboxa{}\fi
  \global\@tempdima\csname tocdim@#2\endcsname
  \leftskip\csname tocleft@#2\endcsname\relax
  \dimen@\csname tocleft@#1\endcsname\relax
  \parindent-\leftskip\advance\parindent\dimen@
  \rightskip\tocleft@pagenum plus 1fil\relax
  \skip@\parfillskip\parfillskip\z@
  \let\numberline\numberline@@sections
  \@nameuse{l@f@#2}%
  \ignorespaces#3\unskip\nobreak\hskip\skip@
  \hb@xt@\rightskip{\hfil\unhbox\@tempboxa}\hskip-\rightskip\hskip\z@skip
  \expandafter\par
  \expandafter\aftergroup\csname tocdim@#2%
  \expandafter\endcsname
  \expandafter\endgroup
              \the\@tempdima\relax
}%
\makeatother

\begin{document}
\title{Brownian motion near an elastic cell membrane: A theoretical study}

\author{Abdallah Daddi-Moussa-Ider}
\email{ider@thphy.uni-duesseldorf.de}
\affiliation
{Biofluid Simulation and Modeling, Fachbereich Physik, Universit\"at Bayreuth, Universit\"{a}tsstra{\ss}e 30, Bayreuth 95440, Germany}

\affiliation
{Institut f\"{u}r Theoretische Physik II: Weiche Materie, Heinrich-Heine-Universit\"{a}t D\"{u}sseldorf, Universit\"{a}tsstra\ss e 1, D\"{u}sseldorf 40225, Germany}

 \author{Stephan Gekle}
    \email{stephan.gekle@uni-bayreuth.de}
    \affiliation
    {Biofluid Simulation and Modeling, Fachbereich Physik, Universit\"at Bayreuth, Universit\"{a}tsstra{\ss}e 30, Bayreuth 95440, Germany}

\begin{abstract}
 Elastic confinements are an important component of many biological systems and dictate the transport properties of suspended particles under flow. 
 In this chapter, we review the Brownian motion of a particle moving in the vicinity of a living cell whose membrane is endowed with a resistance towards shear and bending. 
 The analytical calculations proceed through the computation of the frequency-dependent mobility functions and the application of the fluctuation-dissipation theorem. 
 Elastic interfaces endow the system with memory effects that lead to a long-lived anomalous subdiffusive regime of nearby particles. 
 In the steady limit, the diffusional behavior approaches that near a no-slip hard wall.
 The analytical predictions are validated and supplemented with boundary-integral simulations.
 
\vspace{0.25cm}
{\bfseries Key words:} Anomalous diffusion, cell membrane, singularity methods, biological fluid dynamics.
\end{abstract}

\date{\today}

\maketitle

\tableofcontents

\section{Introduction}

The interactions of nanoparticles with cell membranes play an important role in a variety of biomedical and biotechnological applications.
Prime examples include drug delivery and chemotherapy via nanocarriers~\cite{langer88, naahidi13, al-obaidi15, liu16},  targeted phototherapy~\cite{kirui10, nurunnabi14} and biosensing applications~\cite{xia08, slowing07, suk16}.
During uptake by a living cell via endocytosis~\cite{Doherty_2009, meinel14, AgudoCanalejo_2015}, nanoparticles often come into close vicinity of cell membranes which alter their motion in a complex fashion.

At small length and time scales of motion, the dynamics of the fluid around suspended particles is governed by the steady Stokes equations as long as viscous effects dominate over inertial effects.
In these conditions, a full description of the particles' motion is achieved by the hydrodynamic mobility function which bridges between the particles' velocities and the forces exerted on their surfaces. 
Particle motion in bulk is well understood and has been studied extensively since the pioneering work of Stokes~\cite{stokes51}.
However, in realistic situations, motion often occurs in geometric confinements where the mobility is significantly changed relative to its corresponding bulk value.  
The effect of confining boundaries on nearby particles in a viscous fluid plays an important and crucial role in a variety of technological processes ranging from the rheology of colloidal suspensions~\cite{lowen94, allain95, tanaka00, isa09, glassl10, zimmermann16} to the transport of nanoparticles and various molecules through nanochannels~\cite{stavis05, huh07}.

The first attempt to address the effect of boundaries on the motion of a suspended particle dates back to Lorentz~\cite{lorentz07} who employed the image solution technique to compute the Stokes flow induced by a point-force singularity acting near an infinitely extended planar hard-wall.
This solution technique is applicable when the particle is located at a moderate distance from the wall. 
A fully analytical solution has later been proposed by Brenner~\cite{brenner61} using bispherical coordinates to address the slow motion of a truly finite-sized particle axisymmetrically moving towards a plane surface.
The viscous translational motion parallel to a planar wall has further been investigated using matched asymptotic expansions~\cite{oneill67, goldman67a} finding that the wall introduces a coupling between rotation and translation.
Both the translational and rotational motions have later been reconsidered by Perkins \& Jones who expressed the particle mobility in terms of a set of scalar functions that depend on the sphere radius and distance to a free~\cite{perkins91} or a rigid interface~\cite{perkins92}. 
Lorentz calculations have been extended to account for a finite frequency of motion by Wakiya for the motion parallel to a hard wall~\cite{wakiya64}.

Particle motion through a channel between two adjacent walls has received researchers' attention since a long time ago.
The most simple and intuitive approach to calculating the particle mobility function is due to Oseen~\cite{oseen28} who suggested that the mobility of a sphere confined between two rigid walls could conveniently be approximated by superposition of the leading-order terms from each single wall. 
A more diligent approach has been proposed by \Faxen~\cite{faxen21, faxen22} who computed the particle mobility for the parallel motion to the walls for the particular situations when the particle is in the mid-plane or the quarter-plane between two hard walls~\cite{happel12}.
For an arbitrary position between the two walls, a first-order reflexion theory valid when the particle is far away from the walls has been proposed by Ho \& Leal~\cite{ho74}.
Exact analytical solutions have later been obtained and expressed in terms of a convergent series using the image solution technique for both incompressible~\cite{liron76, bhattacharya02} and compressible flows~\cite{felderhof06twoMem, felderhof10loss, felderhof10echoing}.
For a truly extended particle, multipole expansions~\cite{bhattacharya06, swan10} in addition to joint analytical-numerical solutions have been presented for the perpendicular~\cite{ganatos80a} and parallel motions~\cite{ganatos80b}.
Further theoretical investigations have been carried out near two perpendicular walls~\cite{sano76} and for a sphere confined on the centerline of a rectangular channel~\cite{van10}.

During the past few decades, particle motion in confined geometries regained greater interest after the advent of elaborate experimental techniques which allow an accurate and reliable measurement of the mobility near interfaces.
Among the most efficient techniques that have been utilized are laser~\cite{meyer06, khatibzadeh14} and optical tweezers~\cite{lin00, dufresne01, schaffer07, lele11, traenkle12, mo15, mo15thesis, zahn17}, fluorescence~\cite{kihm04, sadr05} and digital video microscopy~\cite{cui02, eral10, sharma10, benavides12, traenkle13, dettmer14, traenkle16, benavides16}, evanescent wave dynamic light scattering~\cite{mason95b, lobry96, bevan00, clapp01, banerjee05, michailidou09, cichocki10, lisicki12, rogers12, michailidou13, wang14, lisicki14, lisicki15thesis, liu17thesis} and the three-dimensional total internal reflection velocimetry technique~\cite{huang07}.
Analytical calculations of the mobility functions have been extended to include particles near interfaces with partial slip~\cite{lauga05, lauga07, felderhof12}, interfaces separating two mutually immiscible liquids~\cite{bart68} and inside a thin liquid film between two incompressible fluids~\cite{felderhof06film}.
Further, explicit analytical expressions for the flow field induced by a point force acting close to a fluid-fluid interface have been obtained using the image solution technique~\cite{aderogba78}.
For a truly extended particle, analytical solutions have been proposed by Lee and coworkers using a generalization of the method of Lorentz~\cite{lee79} and \enquote{exact} bipolar coordinates~\cite{lee80}.
The effect of small deformations of an initially flat fluid interface on the force and torque experienced by a nearby translating or rotating sphere has also been considered~\cite{berdan81, urzay07}. 
Additional works have been carried out near a viscous interface~\cite{danov95, danov95b, danov98} and an interface covered with surfactant~\cite{shail83b, blawz99a, blawz10theory}.

Unlike fluid-solid or fluid-fluid interfaces, elastic membranes stand apart as they endow the system with memory effects.
Particle motion near a planar fluid membrane endowed with surface tension~\cite{bickel07, bickel14}, bending resistance~\cite{bickel06} or membrane elasticity~\cite{felderhof06, felderhof06b} has been theoretically studied, where it has been found that the steady mobility is universal and identical to that near a no-slip hard wall.
Analytical calculations have be carried out near a realistically modeled elastic red blood cell (RBC) membrane endowed simultaneously with both shear and bending rigidities~\cite{daddi16, daddi16b, daddi16c, daddi17, daddi17d, daddi17e, daddi17b, daddi17c, daddi-thesis} finding that the elasticity of the membrane induces at intermediate time scales an anomalous subdiffusive regime on the nearby particle.
Further theoretical investigations have been recently carried out by Salez and collaborators via thin-film soft lubrication theory~\cite{salez15, saintyves16, rallabandi17}.
Experimentally, particle motion near elastic cell membranes has been investigated using optical traps~\cite{kress05, shlomovitz13, boatwright14, juenger15}, magnetic particle actuation~\cite{irmscher12} and quasi-elastic light scattering~\cite{mizuno00, mizuno04, kimura05} where a significant decrease in the mobility normal to the cell membrane has been observed in line with theoretical predictions.
Additionally, near-membrane dynamics has been used in interfacial microrheological experiments as an efficient and reliable way to extract membrane unknown moduli~\cite{boatwright14, waigh16}.

In this contribution, we provide an overview on the fluid-mediated hydrodynamic interactions and Brownian motion near an elastic cell membrane. 
The analytical predictions proceed through the determination of the Green's function representing the solution of the fluid flow equations due to a point force acting near a membrane.
The frequency-dependent mobility functions of finite-sized particles can then be obtained by combing \Faxen's theorem and multipole expansion.
The mobilities can be expressed in terms of finite series of the ratio between particle radius and membrane distance for the self mobilities, and between radius and interparticle distance for the pair mobilities.
The near-membrane Brownian motion can be addressed using a generalized Langevin formalism and the fluctuation-dissipation theorem.

The reminder of this chapter is organized as follows. 
In Sec.~\ref{sec:membraneMechanics}, we present a relevant model for the membrane and state the traction jump equations stemming from shear and bending deformation modes.
In Sec.~\ref{sec:greensFunctions}, we outline the derivation procedure of the Green's functions associated with a point-force singularity acting near an elastic membrane.
We then review in Sec.~\ref{sec:particleMobility} analytical and numerical results for the particle self- and pair-mobility functions.
In Sec.~\ref{sec:brownianMotion}, we address the Brownian motion of a particle diffusing near a membrane and discuss the diffusional motion parallel and perpendicular to the membrane.
We provide in Sec.~\ref{sec:conclusions} concluding remarks and summaries.


\section{Membrane model}\label{sec:membraneMechanics}

\subsection{Overview}

\begin{figure}  
	\centering
	\def\svgwidth{0.95\linewidth}
	\large
	\includegraphics[scale = 0.45, angle=0]{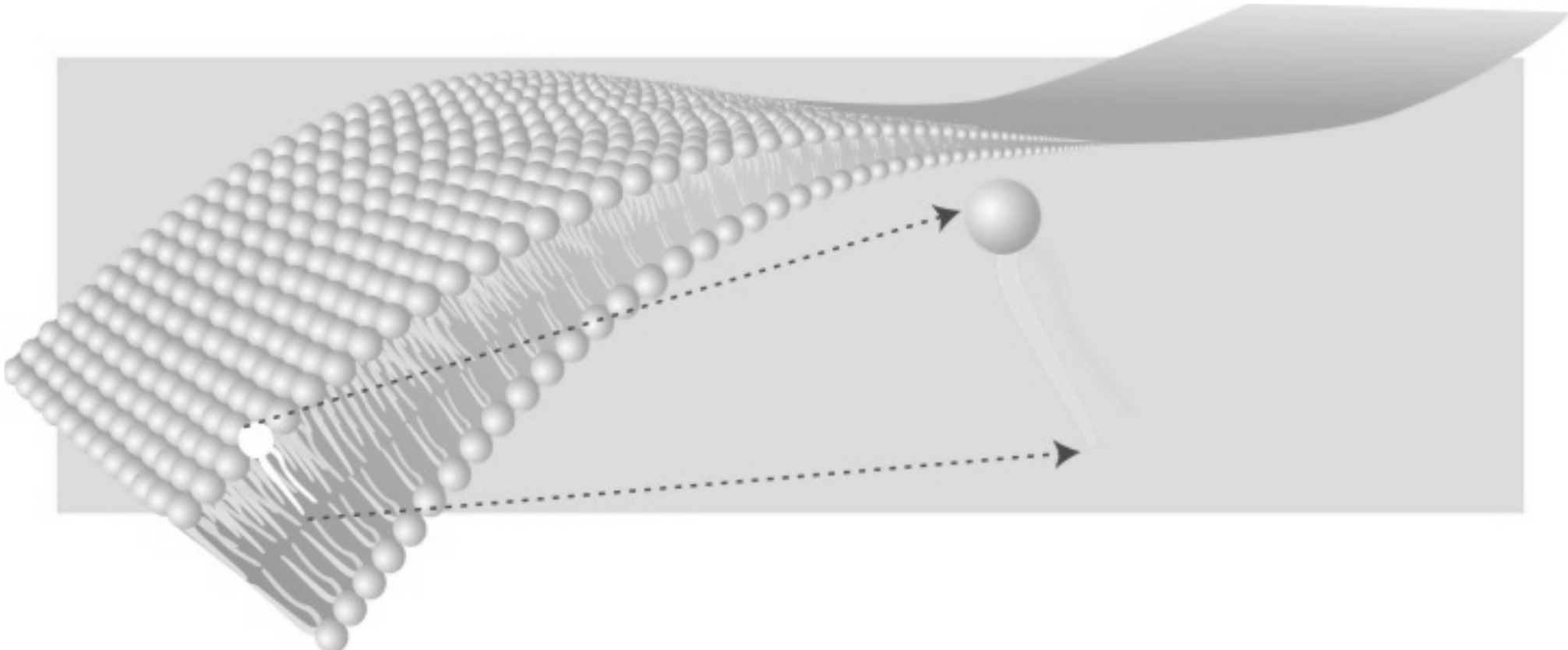}
	\caption{A model elastic membrane consisting of the lipid bilayer.
	 From Peletier and R{\"o}ger, \textit{Arch. Rational Mech. Anal.}, \textbf{193}, 475 (2009), copyright 2009 Springer.}
	\label{lipid}
\end{figure}

The RBC membrane consists of a bilayer composed of amphiphilic lipids that provides the membrane a resistance towards bending.
The latter is well modeled by the Helfrich Hamiltonian~\cite{helfrich73} which is characterized by the bending modulus $\kB$ in addition to the Gaussian curvature modulus $\kappa_\mathrm{K}$~\cite{hu12} (see Fig.~\ref{lipid} for a cartoon schematic of a lipid bilyer.)
Additionally, RBC membranes are endowed with cross-linked cytoskeleton networks that allow the mechanical flexibility required to cope with the shear stresses encountered during cell motion in the microcirculation.
The membrane shear elasticity is commonly described by the Skalak model~\cite{skalak73} which incorporates into a single strain energy functional both the resistance towards shear and area conservation.
The Skalak model is characterized by a shear modulus $\kS$ in addition to an area dilatation (extension) modulus $\kA$~\cite{Lefebvre2007, lac04, krueger11, gekle16, baecher17, schaaf17}.
The latter is typically very large compared to $\kS$ to mimic membrane area incompressibility.
Further elastic models have been proposed in the literature including the neo-Hookean model~\cite{barthes16} and the zero-thickness shell model~\cite{ramanujan98} both of which are characterized by a unique shear modulus~$\kS$.

\subsection{Membrane parametrization}

The membrane is modeled as a zero-thickness, two-dimensional elastic sheet that is infinitely extended along the plane $z=0$ in the absence of external loads.
The presence of a nearby particle creates an imbalance in the stress tensor across the interface which leads to the membrane deformation.

We adopt a local coordinate system $\{x, y\}$ for the undeformed membrane surface and describe a mapping that assigns each pair $(x,y)$ to the position vector $\vect{A}$ defined in the Cartesian coordinate system as~\cite{deserno15}
\begin{equation}
  \vect{A} (x,y) = x \eX + y \eY \, .
\end{equation}

After deformation, the position vector reads
\begin{equation}
    \vect{a} (x,y) = (x + u_x) \eX + (y + u_y) \eY + u_z \eZ \, ,
\end{equation}
where $u_x$, $u_y$ and $u_z$ are the Cartesian components of the displacement vector $\vect{u} (x,y)$.
Hereafter, we will use capital and small roman letters for the undeformed and deformed states, respectively.
The membrane can be defined by the covariant base vectors 
\begin{equation}
  \vect{g}_1 :=  \frac{\partial \vect{a}}{\partial x} \, , \qquad 
  \vect{g}_2 :=  \frac{\partial \vect{a}}{\partial y} \, , 
  \label{covariantBasis}
\end{equation}
which are tangent to the membrane surface.
The unit normal vector reads
\begin{equation}
  \vect{n} = \frac{\gOne \times \gTwo}{|\gOne \times \gTwo|}  \, .
\end{equation}

The covariant components of the metric tensor are defined by the scalar product $g_{\alpha\beta} = \vect{g}_{\alpha} \cdot \vect{g}_{\beta}$.
The contravariant tensor $g^{\alpha\beta}$ (conjugate metric) is the inverse of the metric tensor.

\subsection{Traction jump equations}

Depending on the biological composition of the cell, the membrane may exhibit a resistance towards shear and/or bending.
Hereafter, we will state the traction jump equations stemming from the membrane shear and bending resistances. 

\subsubsection{Shear}

In order to account for the shear deformation mode, we introduce the in-plane transformation gradient~\cite{ogden97}
\begin{equation}
	\Intd {a}_\alpha = {F}_{\alpha\beta} \, \Intd {A}_\beta \, , 
\end{equation}
bridging between the infinitesimal displacements in the deformed and undeformed spaces.
Eq.~\eqref{covariantBasis} leads to an expression of the transformation gradient tensor in term of the dyadic product $\vect{F} =  \vect{g}_\alpha \otimes \vect{G}^\alpha$ .

We now define the right Cauchy-Green deformation tensor $C_{\alpha\beta} = F_{\gamma\alpha} F_{\gamma\beta}$ whose invariants are given by Green and Adkins as~\cite{green60, zhu14}
\begin{subequations}
\begin{align}
I_1 &= G^{\alpha\beta}  g_{\alpha\beta} - 2 \, , \\
I_2 &= \det G^{\alpha\beta}  \det g_{\alpha\beta} - 1 \, .
\end{align}
\end{subequations}

Provided knowledge of the membrane constitutive law, the contravariant components of the stress tensor $\tau^{\alpha\beta}$ can readily be obtained from~\cite{lac04}
\begin{equation}
\tau^{\alpha\beta} = \frac{2}{\JS} \frac{\partial W}{\partial I_1} \, G^{\alpha\beta} + 2\JS \, \frac{\partial W}{\partial I_2} \, g^{\alpha\beta} \, ,
\label{stressTensor}
\end{equation}
wherein $W$ is the areal strain energy functional and $\JS := \sqrt{1+I_2}$ is the Jacobian determinant, representing  the  ratio  between  the  deformed  and
undeformed local surface areas. 

The tractions jumps due to shear can be obtained by writing the equilibrium equations balancing between the membrane elastic forces and external loads, as it has been detailed in Ref.~\onlinecite{daddi16}.

\subsubsection{Bending}

According to the Helfrich model, the bending energy for a flat sheet is described by a quadratic curvature-elastic continuum model as~\cite{Guckenberger_preprint}
\begin{equation}
	E_\mathrm{B} = \int_S 2\kB H^2 \, \Intd S + \int_S \kappa_\mathrm{K} K \, \Intd S \, ,  \label{HelfrichHamiltonian}
\end{equation} 
where $H$ and $K$ are the mean and Gaussian curvatures, respectively. 
The traction jump equations across a membrane endowed with a resistance towards bending can readily be obtained via a variational approach by minimizing the sum of the bending and external potential energy to obtain~\cite{guckenberger17}
\begin{equation}
\Delta \vect{f}        = -2\kB \left( 2(H^2-K) + \Delta_\parallel \right) H \, \vect{n} \, ,   \label{tractionJumpHelfrich}
\end{equation}
and it is commonly denominated the Euler-Lagrange equation~\cite{jenkins77, powers10, laadhari10, sinha15}.
The mean and Gaussian curvatures are respectively expressed by
\begin{equation}
H = \frac{1}{2} \, b_\alpha^\alpha \, , \qquad 
K = \mathrm{det~} b_\alpha^\beta \, , 
\end{equation}
with $b_\alpha^\beta$ being the mixed version of the curvature tensor.

The resulting linearized traction jumps across a membrane endowed with both shear and bending rigidities read \cite{daddi16}
\begin{align}
	\Delta f^{\beta} &= -\frac{\kappa_\mathrm{S}}{3} \big( \Delta_{\parallel} u_\beta + (1+2C) e_{,\beta} \big)  \, , \quad \beta \in \{x,y\} \, ,  \label{tangentialTractionJump_Shear} \\
 	\Delta f^{z} &= \kappa_\mathrm{B}   \Delta_{\parallel}^2 u_z \, , \label{normalTractionJump_Bending}
\end{align}
where  $\Delta_\parallel f = f_{,xx}+f_{,yy}$ is the Laplace-Beltrami of a given function $f$. 
It can clearly be seen that at leading order in deformation, shear does not introduce a normal traction jump. 
Additionally, bending as derived from the Helfrich model does not introduce a jump in the tangential traction~\cite{guckenberger17}.

Having introduced a model for the membrane and stated the underlaying traction jumps, we present in the next section the methodology for computing the Green's functions associated with a point force acting near a membrane.


\section{Green's functions}\label{sec:greensFunctions}

\subsection{Mathematical formulation}

In the Stokes regime, the fluid flow around a particle obeys the steady Stokes equations~\cite{kim13}
\begin{align}
\eta \bNabla^2 \bv(\R) - \bNabla p(\R) + \vect{F} \delta(\R-\R_0) &= 0 \, , \label{stokesEq_1} \\
\bNabla \cdot \bv (\R) &= 0 \, ,  \label{stokesEq_2}
\end{align}
where the fluid velocity $\vect{v}$ and pressure field $p$ are due to a concentrated point force $\vect{F} \delta(\R-\R_0 )$ induced by a point-particle immersed at the position~$\R_0$.
The solution of these equations for the velocity are expressed in terms of the Green's function
\begin{equation}
	v_\alpha (\R) = \G_{\alpha\beta} (\R,\R_0) F_\beta \, ,  
\end{equation}
which in an unbounded and infinite fluid is commonly known as the Oseen tensor,
\begin{equation}
	\G_{\alpha\beta}^{\mathrm{O}} (\R,\R_0) = \frac{1}{8\pi\eta} 
	\left( \frac{\delta_{\alpha\beta}}{s} + \frac{s_\alpha s_\beta}{s^3} \right) \, , 
	\label{oseenTensor}
\end{equation}
wherein $\vect{s} := \R-\R_0$ and $s := |\vect{s}|$.
The pressure field is expressed as $p = \PP_\beta F_\beta$ where $\PP_\beta = s_\beta/(4\pi s^3)$. 

The presence of an elastic membrane introduces a correction to the Green's function that is dependent on the point-force position and the actuation frequency of the system, as it is explained below.

\subsection{Fourier transform technique}

 \begin{figure}
  \begin{center}
	  \includegraphics[scale=0.95]{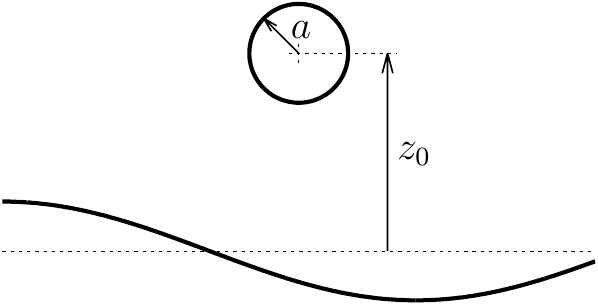}
  \end{center}
  \caption{Illustration of particle motion near an elastic membrane. 
The membrane is extended in the $xy$ plane and the solid particle of radius~$a$ is located above the membrane at $z=z_0$.}
  \label{Illustration_Planar}
  \end{figure}

The Stokes equation near a planar membrane are more conveniently solved using 2D Fourier transform technique.
The latter consists of transforming the partial differential equations~\eqref{stokesEq_1} and~\eqref{stokesEq_2} governing the fluid motion into ordinary differential equations for the out-of-plane coordinate~$z$.

We now assume that the point-force $\F$ is acting at the position $\vect{r}_0 = (0,0, z_0)$ with $z_0 > 0$, above an elastic membrane infinitely extended in the plane $z=0$ (see Fig.~\ref{Illustration_Planar} for an illustration of the system setup.)
Without loss of generality, we assume that the fluid has the same and constant dynamic viscosity $\eta$ on both sides of the membrane.

We define the spacial 2D (forward) Fourier transform
\begin{equation}
\mathscr{F} \{ f(\boldsymbol{\rho}) \} = \tilde{f}(\vect{q}) = 
\int_{\mathbb{R}^2} f(\boldsymbol{\rho}) e^{-i \vect{q} . \boldsymbol{\rho}} \, \Intd^2 \boldsymbol{\rho} \, , 
\label{2DFourierTransform}
\end{equation}
together with its inverse transform
\begin{equation}
\mathscr{F} ^{-1} \{ \tilde{f} (\vect{q}) \} = f(\boldsymbol{\rho}) =
\frac{1}{(2 \pi)^2}
\int_{\mathbb{R}^2} \tilde{f} (\vect{q}) e^{i \vect{q} . \boldsymbol{\rho}} \, \Intd^2 \vect{q} \, , 
\label{inverse2DFourierTransform}
\end{equation}
where  $\boldsymbol{\rho} = (x,y)$ is the projection of the position vector $\vect{r}$ onto the $xy$ plane, and $\vect{q} = (q_x, q_y)$ is the Fourier transform variable.

It turns out to be convenient to employ the orthogonal coordinate system previously introduced by Bickel~\cite{bickel07, bickel06} in which all the vector fields are decomposed into longitudinal, transverse and normal components. 
For a given Fourier transformed quantity~$\tilde{\vect{A}}$, whose horizontal components in the Cartesian coordinate basis are $(\tilde{A}_x, \tilde{A}_y)$, its components in the new orthogonal base are given by the following orthogonal transformation
\begin{equation}  \label{transformation}
\left( 
      \begin{array}{c}
      \tilde{A}_x \\
      \tilde{A}_y 
      \end{array}
\right)
=
\frac{1}{q}
\left( 
      \begin{array}{cc}
      q_x & q_y\\
      q_y & -q_x
      \end{array}
\right)
\left( 
      \begin{array}{c}
      \tilde{A}_l \\
      \tilde{A}_t 
      \end{array}
\right) \, ,
\end{equation}
where $\tilde{A}_l$ and $\tilde{A}_t$ refer to the longitudinal and transverse components, respectively, and $q := |\vect{q}|$ is the wavenumber.
Clearly, the normal component $\tilde{A}_z$ is left unchanged by this transformation.
We note that the inverse transformation is also given by the same transformation matrix in Eq.~\eqref{transformation}.

As the membrane shape depends on the history of particle motion, we will perform further a Fourier analysis in time.
For a function $f(t)$ expressed in the temporal space, its (forward) Fourier transform to the frequency domain is defined as
\begin{equation}
f(\omega) = \int_{\mathbb{R}} f(t) e^{-i \omega t} \, \Intd t \, , \label{fourierInTime}
\end{equation}
and the inverse Fourier transform back to the real space reads
\begin{equation}
f(t) = \frac{1}{2 \pi} \int_{\mathbb{R}} f ( \omega ) e^{i \omega t} \, \Intd \omega \, .
\end{equation}
Since both the spatial and temporal Fourier transforms are performed here, we will adopt the convention where the two functions $f(t)$ and $f(\omega)$ are distinguished only by their argument.
The tilde will therefore be reserved for the spatial 2D Fourier transform.

The projected Stokes equations~\eqref{stokesEq_1} and~\eqref{stokesEq_2} upon 2D Fourier transformation result in four ordinary differential equations with the variable $z$~\cite{bickel07}.
The components of the Green tensor in Fourier space are~\cite{daddi16}
\begin{equation}
\left( 
      \begin{array}{c}
      \vt_z\\
      \vt_l\\
      \tilde{v_t}
      \end{array}
\right)
=
\left( 
      \begin{array}{ccc}
      \tilde{\mathcal{G}}_{zz} & \tilde{\mathcal{G}}_{zl} & 0\\
      \tilde{\mathcal{G}}_{lz} & \tilde{\mathcal{G}}_{ll} & 0\\
      0 & 0 & \tilde{\mathcal{G}}_{tt}
      \end{array}
\right)
\left( 
      \begin{array}{c}
      F_z \\
      F_l \\
      F_t
      \end{array}
\right) \, .
\label{greenFunctions}
\end{equation}

The governing equations are subject to the boundary conditions at the undisplaced membrane, namely (a)~the natural continuity of the velocity components and (b)~the discontinuity of the fluid stress tensor caused by membrane shear and bending described in the previous section.
Technical details regarding the derivation of the Green's functions have been discussed in Refs.~\onlinecite{daddi16} and \onlinecite{daddi17}.

For future reference, we define at this point the characteristic frequency for shear and bending, respectively by
\begin{equation}
\beta = \frac{12 z_0 \eta\omega}{(1+C)\kS} \, , \qquad
\betaB = 2z_0 \left( \frac{4\eta\omega}{\kB} \right)^{1/3} \, .
\end{equation}

Moreover, we define the reduced bending modulus, quantifying the coupling between the shear and bending deformation modes as $\EB = \kB/(\kS z_0^2)$.

In the vanishing-frequency limit, or equivalently for infinite membrane shear and bending rigidities, the Green tensor near a hard wall is recovered~\cite{vonHansen11}
\begin{equation}
\Gmatr^\mathrm{B} (\R) = \Gmatr^{\mathrm{O}} (\vect{s}) - \Gmatr^{\mathrm{O}} (\Rbig) + \Gmatr^{\mathrm{D}} (\Rbig) - \Gmatr^\mathrm{SD} (\Rbig) \, ,
\end{equation}
and it is commonly denominated the Blake tensor~\cite{blake71}.
Here $\Rbig:=\R - \overline{\R_0}$ with $\overline{\R_0} = (0,0,-z_0)$ being the position of the Stokeslet image and we recall that $\vect{s} := \R-\R_0$.
Furthermore, $r:= |\R|$ and $R:=|\Rbig|$.
Here $\Gmatr^{\mathrm{D}}$ is the force (Stokeslet) dipole and $\Gmatr^\mathrm{SD}$ is the source dipole, respectively given by
\begin{align}
\G^{\mathrm{D}}_{\alpha\beta} (\Rbig) &= \frac{2z_0^2 (1-2\delta_{\beta z})}{8\pi\eta} 
		    \left( \frac{\delta_{\alpha\beta}}{R^3} - \frac{3R_\alpha R_\beta}{R^5} \right) \, , \\
\G^\mathrm{SD}_{\alpha\beta} (\Rbig) &= \frac{2z_0 (1-2\delta_{\beta z})}{8\pi\eta}
		    \bigg(  
		    \frac{\delta_{\alpha\beta} R_z}{R^3} - \frac{\delta_{\alpha z} R_\beta}{R^3} \notag \\
		    &+ \frac{\delta_{\beta z} R_\alpha}{R^3}
		    -\frac{3R_\alpha R_\beta R_z}{R^5}
		    \bigg) \, .
\end{align}

It is worth noting that near a single membrane, shear and bending present a decoupled nature.
As a result, the Green's functions near a membrane endowed simultaneously with both shear and bending rigidities can appropriately be determined by linear superposition of the Green's functions associated with membranes with pure shear and pure bending as obtained independently.
However, this interesting feature is not observed near two parallel elastic membranes~\cite{daddi16b} or curved membranes~\cite{daddi17d, daddi17e, daddi17b, daddi17c} where a strong coupling behavior occurs.

The near-membrane Green's functions serve as basis for the determination of the effect of elastic membranes on the motion of suspended particles.
Particularly, we will be interested in the particle self- and pair-mobility functions.


\section{Particle hydrodynamic mobility}\label{sec:particleMobility}

 \begin{figure}
  \begin{center}
	  \includegraphics[scale=0.4]{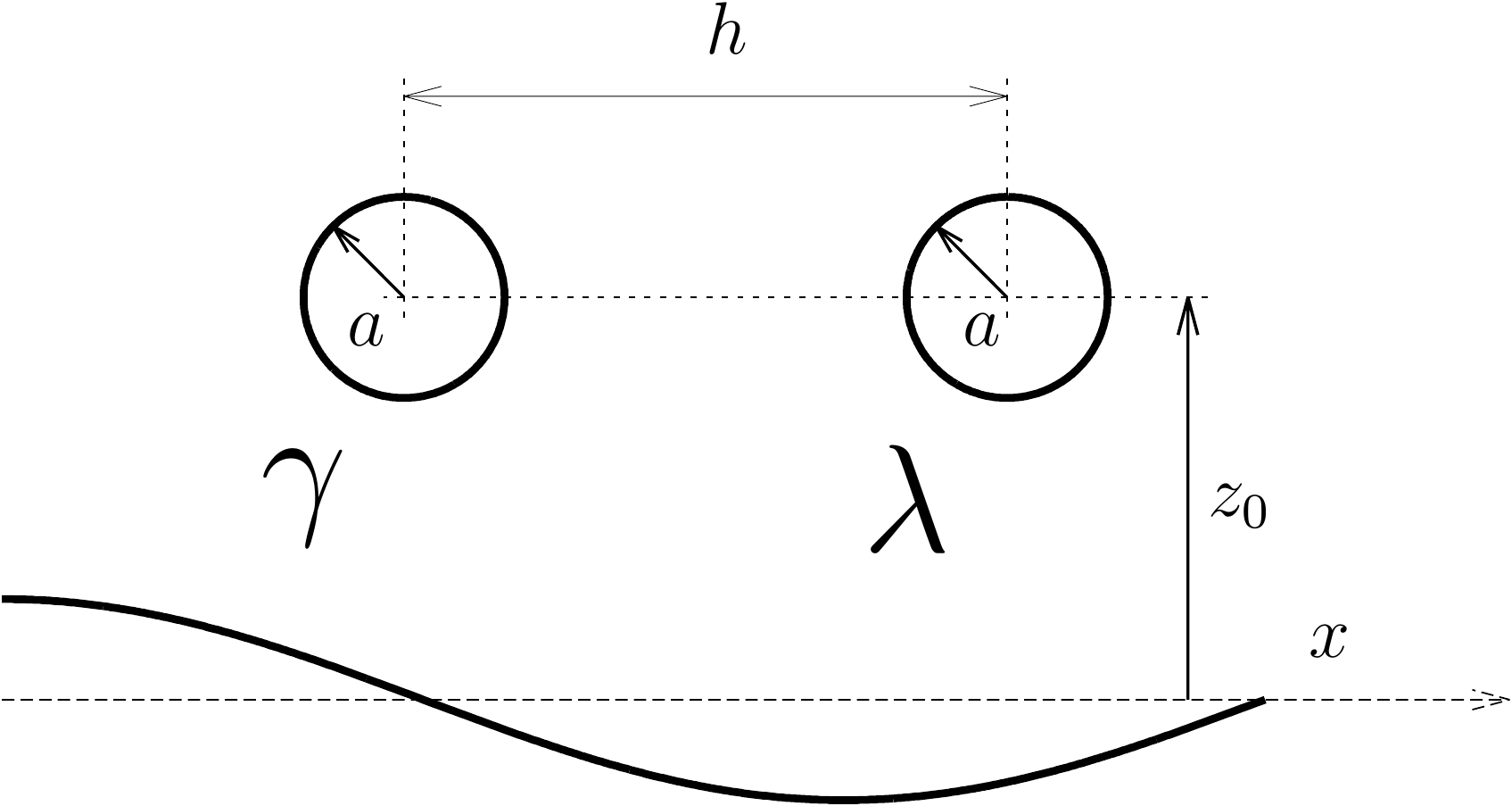}
  \end{center}
  \caption{Illustration of a pair of particles near a planar elastic membrane. 
  The particles labeled $\gamma$ and $\lambda$ have the same radius~$a$ and are located at $\R_{\gamma}=(0,0,z_0)$ and $\R_{\lambda} = (h,0,z_0)$, respectively. }
  \label{Illustration_Pair}
  \end{figure}

We consider a representative configuration of a pair of particles denoted $\gamma$ and $\lambda$ located a distance $z_0$ above a planar elastic membrane and a distance $h$ apart, as schematically sketched in Fig.~\ref{Illustration_Pair}.
Assuming a force density $\vect{f}_\lambda$ acting at the particle $\lambda$ located at $\R_\lambda$, the disturbance velocity field at $\R$ can be written as 
\begin{equation}
  \bv (\R, \R_\lambda, \omega) = \bv^{\mathrm{S}} (\R, \R_\lambda) + \bv^* (\R, \R_\lambda, \omega) \, , \label{fluidVelocitySplitUp}
\end{equation}
where a Fourier transformation has been applied to the temporal dependence of all fields.
Here $\bv^{\mathrm{S}}$ denotes the induced fluid flow in an unbounded and infinite fluid and $\bv^*$ is the flow field required to satisfy the boundary conditions at the membrane, also known as the reflected flow field.
The disturbance field can be written as an integral over the surface of the sphere $\lambda$ as
\begin{equation}
  \bv (\R, \R_{\lambda}, \omega) = \oint_{S_{\lambda}} \Gmatr (\R, \R', \omega) \cdot {\vect{f}_\lambda} (\R', \omega) \, \Intd^2 \R' \, ,
\label{fluidVelocityIntPointForce}
\end{equation}
where $\Gmatr $ is the Green's function associated with a point-force acting at a point located at $\R'$ belonging to the surface of the particle~$\lambda$.
Similar, the Green's functions can be split up into two distinct contributions,
\begin{equation}
\Gmatr (\R, \R', \omega) = \Gmatr^{\mathrm{O}} (\R, \R') + \Gmatr^\mathrm{M} (\R, \R', \omega) \, ,
\label{eqn:defDeltaG}
\end{equation}
where again $\Gmatr^{\mathrm{O}}$ is the Oseen tensor given by Eq.~\eqref{oseenTensor} and $\Gmatr^\mathrm{M}$ represents the frequency-dependent correction to the Green's function due to the presence of the membrane.

Far away from the particle~$\lambda$, the integration vector variable $\R'$ in Eq.~\eqref{fluidVelocityIntPointForce} can conveniently be expanded around the particle center $\R_{\lambda}$ following a multipole expansion approach. 
Up to the second order and assuming a constant force density over the particle surface, the disturbance velocity can be approximated by~\cite{swan07, aponte16}
\begin{equation}
  \begin{split}
    \bv (\R, \R_{\lambda},\omega) \approx \left( 1+\frac{a^2}{6} \boldNabla_{\R_{\lambda}}^2  \right) \Gmatr (\R, \R_{\lambda}, \omega) \cdot {\vect{F}} (\omega) \, , 
  \end{split}
\label{flowField_MultipoleExp}
\end{equation}
where $\boldNabla_{\R_{\lambda}}$ stands for the gradient operator taken with respect to the singularity position $\R_\lambda$.
For a single sphere in bulk, the flow field given by Eq.~\eqref{flowField_MultipoleExp} satisfies exactly the no-slip boundary conditions at the surface of the sphere~\cite{kim06}.

Using \Faxen's theorem, the translational velocity of the adjacent particle~$\gamma$ in this flow reads~\cite{swan07, aponte16}
\begin{equation}
\vect{V}_{\gamma}(\omega)      = \mu_0 {\vect{F}_{\gamma}} (\omega) + \left( 1+\frac{a^2}{6} \boldNabla_{\R_{\gamma}}^2 \right) \bv (\R_{\gamma}, \R_{\lambda}, \omega)  \, , \label{Faxen_trans}
\end{equation}
where $\mu_0 := 1/(6\pi\eta a)$ denotes the usual translational bulk mobility given by the Stokes law.
We further emphasize that the disturbance flow $\bv$ incorporates both the disturbance from the particle~$\lambda$ and the disturbance caused by the presence of the membrane.
By inserting Eq.~\eqref{flowField_MultipoleExp} into \Faxen's formula stated by Eqs.~\eqref{Faxen_trans}, the frequency-dependent translational tensor can be obtained from the \emph{total} Green's functions as~\cite{daddi16c}
\begin{equation}
		    \mi^{\gamma\lambda} (\omega) = \left( 1+\frac{a^2}{6} \boldNabla_{\R_{\gamma}}^2 \right) 
		      \left( 1+\frac{a^2}{6} \boldNabla_{\R_{\lambda}}^2  \right) \Gmatr (\R_{\gamma}, \R_{\lambda}, \omega)  \, . \label{particlePairMobility_tran} 
\end{equation}
For the self-mobilities, only the correction to the flow field $\bv^*$ due to the presence of the membrane in Eq.~\eqref{fluidVelocitySplitUp} should be considered in \Faxen's formulas. 
Therefore, the frequency-dependent self-mobility tensors are directly determined from the \emph{correction} to the Green's functions to obtain~\cite{daddi16c}
\begin{equation}
	 \begin{split}
	  \mi^{\gamma\gamma} (\omega) &= \mu_0 \, \mathbf{1}  + \lim_{\R \to \R_\gamma} \left( 1+\frac{a^2}{6} \boldNabla_{\R}^2 \right) \\
	 		       &\times \left( 1+\frac{a^2}{6} \boldNabla_{\R_\gamma}^2  \right) \Gmatr^\mathrm{M} (\R, \R_\gamma, \omega)  \, , \label{particleSelfMobility_tran} 
	 \end{split}		
\end{equation}
where $\mathbf{1}$ denotes the unit tensor.
The self- and pair-mobilities in the often-used point-particle approximation are readily obtained by taking the vanishing radius limit in Eqs.~\eqref{particlePairMobility_tran} and~\eqref{particleSelfMobility_tran}, respectively.

\subsection{Self-mobility functions}

We denote by $\mi^{\gamma\gamma}  = \mi^{\mathrm{S}}$ the self-mobility tensor which can conveniently be expressed in terms of power series of $\epsilon := a/z_0$. 
The self-mobility tensor near a planar membrane has the form
\begin{equation}
    \mi^{\mathrm{S}} = 
    \left( 
    \begin{array}{ccc}
    \mu_{xx}^{\mathrm{S}} & 0                        & 0 \\
    0                        & \mu_{yy}^{\mathrm{S}} & 0 \\
    0 & 0 & \mu_{zz}^{\mathrm{S}} \\
    \end{array}
    \right) \, , 
\end{equation}
where the $xx$ and $yy$ components of the self mobility are equal since they are both associated with a motion parallel to the membrane.
The $zz$ component is associated with the axisymmetric motion perpendicular to the membrane.

For a planar membrane, the total mobility correction is obtained by linear superposition of the individual contributions due to shear and bending.
Analytical expressions can be found in Ref.~\onlinecite{daddi16c}.
The known mobility corrections near a hard wall with stick boundary conditions are recovered in the vanishing frequency limits, namely
\begin{equation}
\lim_{\beta,\betaB \to 0} \frac{\Delta \mu_{zz}^{\mathrm{S}}}{\mu_0} = -\frac{9}{8} \, \epsilon + \frac{1}{2} \, \epsilon^3 -\frac{1}{8} \, \epsilon^5  \, .
\label{hardWallPerp}
\end{equation}
for the perpendicular motion, and 
\begin{equation}
\lim_{\beta,\betaB \to 0} \frac{\Delta \mu_{xx}^{\mathrm{S}}}{\mu_0} = -\frac{9}{16} \, \epsilon + \frac{1}{8} \, \epsilon^3 -\frac{1}{16} \, \epsilon^5  \, ,
\label{hardWallPara}
\end{equation}
For the motion parallel to the membrane~\cite{faxen22, swan10}.

\begin{figure}
\begin{center}
   \includegraphics[scale = 0.45]{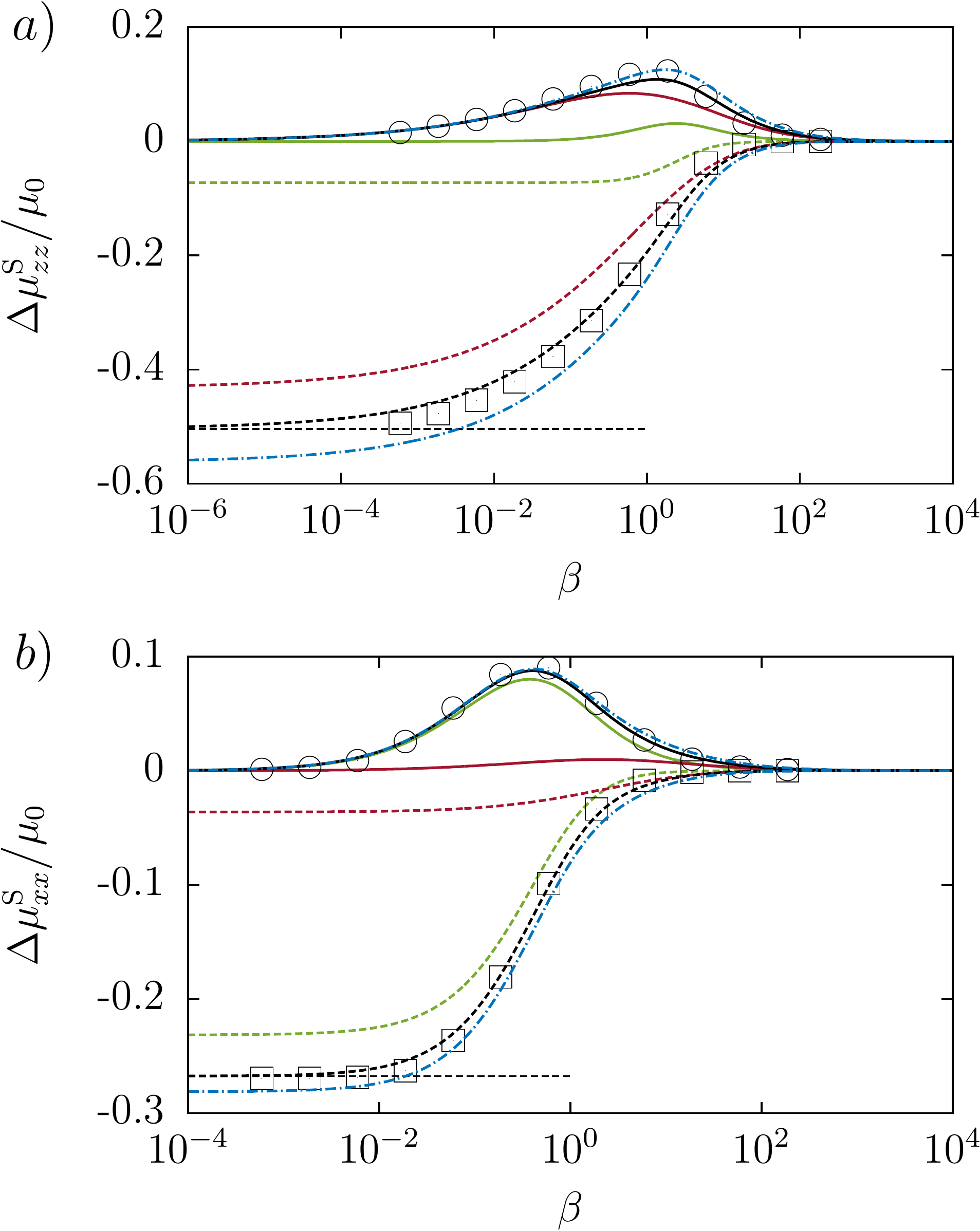}
   \caption{ (Color online) 
   The scaled frequency-dependent correction to the particle self-mobility versus the scaled frequency for motion perpendicular $(a)$ and parallel $(b)$ to the membrane.
   The particle is set at $z_0 = 2a$. 
   The membrane parameters are $z_0^2 \kS / \kB = 3/2$ and $C=1$.
   The analytical predictions are presented as dashed lines for the real part, 
   and as solid lines for the imaginary part.
   Symbols correspond to BIM simulations. 
   The shear- and bending-related contributions are shown in green and red, respectively.
   The dotted-dashed line shown in blue corresponds to the leading-order correction in the particle self-mobility, determined in Ref.~\cite{daddi16}.
   Horizontal dashed lines represent the mobility corrections near a hard-wall as stated by Eqs.~\eqref{hardWallPerp} and~\eqref{hardWallPara}. 
   From Daddi-Moussa-Ider and Gekle, \textit{J.~Chem. Phys.}, {\bf 145}, 014905 (2016), copyright 2016 American Institute of Physics.
   }
   \label{mobiCorrSelf}
\end{center}
\end{figure}

In Fig.~\ref{mobiCorrSelf} $a)$, we show the particle scaled self-mobility corrections versus the scaled frequency $\beta$, for a particle located above a membrane at $z_0 = 2a$.
We consider a reduced bending modulus $\EB=2/3$ corresponding to $\betaB = 2 (\beta/B)^{1/3}$ for which shear and bending manifest themselves equally~\cite{daddi16c}. 
We observe that the real part is a monotonically increasing function with respect to frequency while the imaginary part exhibits the typical bell-shaped dependence on frequency.
In the limit of an infinite actuation frequency, both the real and imaginary parts of the mobility correction vanish, and thus one recovers the behavior in bulk fluid.
For the perpendicular motion, we observe that the particle mobility correction is primarily determined by the bending-related part and that shear does not play a major role.

In order to assess the appropriateness and accuracy of our analytical predictions, computer simulations based on the boundary integral method (BIM) \cite{PozrikidisBook92, pozrikidis01} have been performed.
Technical details regarding the numerical implementation can be found in Refs.~\onlinecite{daddi16b,guckenberger16}. 
A very good agreement has been obtained between the analytical predictions and numerical simulations over the whole range of applied frequencies. 
Additionally, the validity of the point-particle approximation employed in Ref.~\cite{daddi16} is probed. 
While this approximation somehow underestimates the particle mobilities, it surprisingly can lead to a good prediction, even though  the particle is set only one diameter above the membrane.

The mobility corrections in the parallel direction are shown in Fig.~\ref{mobiCorrSelf} $b)$. We observe that the total correction is mainly determined by the shear-related part in contrast to the perpendicular case where bending effect dominates.

\subsection{Pair-mobility functions}

We denote by $\mi^{ab, \gamma\lambda}  = \mi^{ab, \mathrm{P}}$ the pair-mobility tensor which can be expressed as a power series of $\sigma:= a/h$. 
For the present configuration, we have
\begin{equation}
   \mi^{\mathrm{P}} = 
   \left( 
   \begin{array}{ccc}
   \mu_{xx}^{\mathrm{P}} & 0                        & \mu_{xz}^{\mathrm{P}} \\
   0                        & \mu_{yy}^{\mathrm{P}} & 0 \\
   \mu_{zx}^{\mathrm{P}} & 0 & \mu_{zz}^{\mathrm{P}} \\
   \end{array}
   \right)  \, .
\end{equation}
The off-diagonal components $xz$ and $zx$ of the pair-mobility have same absolute value and differ only in sign.
We further note that $\mu_{xz}^{tr} = \mu_{zx}^{rt}$ as required by the symmetry of the mobility tensors.

In an unbounded geometry, the bulk pair mobilities for the motion perpendicular to and along the line of centers can be obtained from the Oseen tensor as~\cite[p. 190]{kim05}
\begin{equation}
 \frac{ \mu_{yy}^{\mathrm{P}}}{\mu_0} = \frac{ \mu_{zz}^{\mathrm{P}}}{\mu_0} = \frac{3}{4} \, \sigma + \frac{1}{2} \, \sigma^3  \, , \quad 
 \frac{ \mu_{xx}^{\mathrm{P}}}{\mu_0} = \frac{3}{2} \, \sigma - \sigma^3  \, ,
\end{equation}
and $\mu_{xz}^{\mathrm{P}} = \mu_{zx}^{\mathrm{P}} = 0$.
Physically, the parameter $\sigma$ only takes values between 0 and $1/2$ as overlap between the two particles is not allowed.


\begin{figure*}
  \begin{center}
     \includegraphics[scale = 0.75]{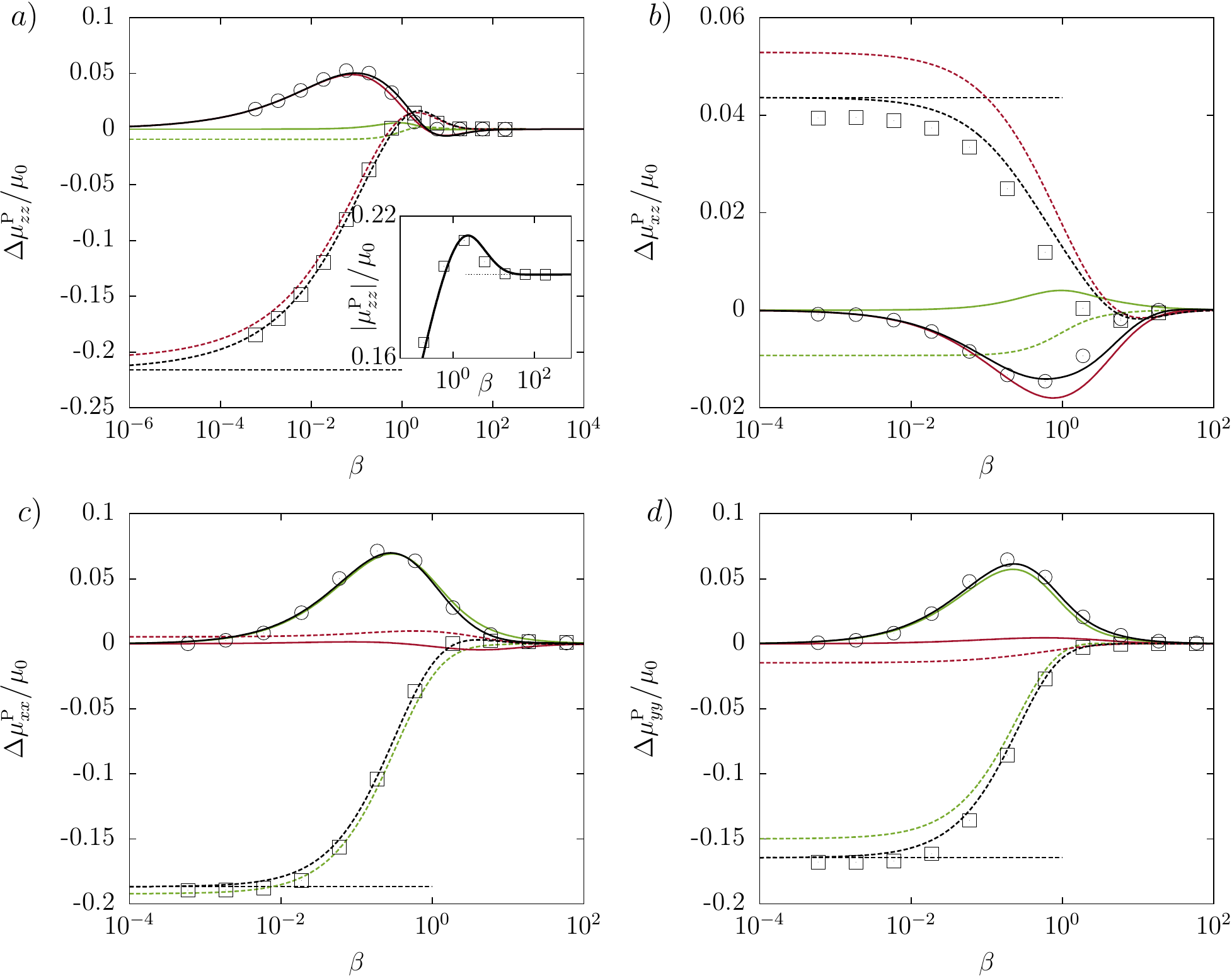}
     \caption{ (Color online) 
     The scaled corrections to the pair-mobility versus the scaled frequency.
     The two particles are located above the membrane at $z_0 = 2a$ with an interparticle distance $h=4a$. 
     The color code is the same as in Fig.~\ref{mobiCorrSelf}.
     The inset in $a)$ shows that the amplitude of the \emph{total} pair-mobility component $zz$ even exceeds its bulk value (dotted line) in a frequency range around $\beta\sim 1$.
     From Daddi-Moussa-Ider and Gekle, \textit{J.~Chem. Phys.}, {\bf 145}, 014905 (2016), copyright 2016 American Institute of Physics. }
     \label{mobiCorr}
  \end{center}
\end{figure*}

Near an elastic membrane, the corrections to the pair-mobilities can readily be computed from Eq.~\eqref{particlePairMobility_tran} and are conveniently expressed in terms of convergent infinite integrals \cite{daddi16c}.
In the vanishing frequency limit, the pair-mobilities near a hard-wall with stick boundary conditions are recovered, as first computed by Swan and Brady \cite{swan07}.


In Fig.~\ref{mobiCorr}, we plot the particle pair mobilities as functions of the dimensionless frequency $\beta$ for $z_0=2a$ and $h=4a$.
We observe that the real and imaginary parts have typically the same evolution as the self mobilities. 
The membrane bending manifests itself in a more pronounced way for the components $zz$ and $xz$ whereas shear effect dominates for the components $xx$ and $yy$.
However, we observe two remarkably different effects:
Firstly, the amplitude of the pair-mobility normal-normal component $zz$ even exceeds its bulk value in a small frequency range.
This enhanced mobility has not been observed for the self mobilities and may result in a short-lasting superdiffusive behavior~\cite{daddi16c}.
Secondly, considering the components $xx$ and $xz$ we find that, unlike the self mobilities, shear and bending may have opposite contributions to the total pair mobilities. For the $xz$ component this implies the interesting behavior that hydrodynamic interactions can be either attractive or repulsive, depending on the membrane properties and the geometric configuration of the pair of particles. 
This behavior will be investigated in more detail in the next subsection.

For future reference, we note that each component of the frequency-dependent particle self- and pair-mobility tensor can conveniently be cast in the form~\cite{daddi16c}
\begin{equation}
 \frac{\mu (\omega)}{\mu_0} = b + \int_{0}^{\infty}\frac{\varphi_1 (u)}{\varphi_2 (u) + i\omega T } \, \Intd u \, , 
 \label{eqn:muGeneral}
\end{equation}
where indices and superscripts have been omitted for the sake of readability.
Here $b$ denotes the scaled bulk mobility, and the integral term represents either the shear- or bending-related parts in the mobility correction.
Note that $\varphi_1$ and $\varphi_2$ are real functions which do not depend on frequency. Moreover, $\varphi_2 \ge 0$ in the integration range over~$u$.

\subsection{Perpendicular steady motion}

\begin{figure}
  \begin{center}
     \includegraphics[scale = 0.45]{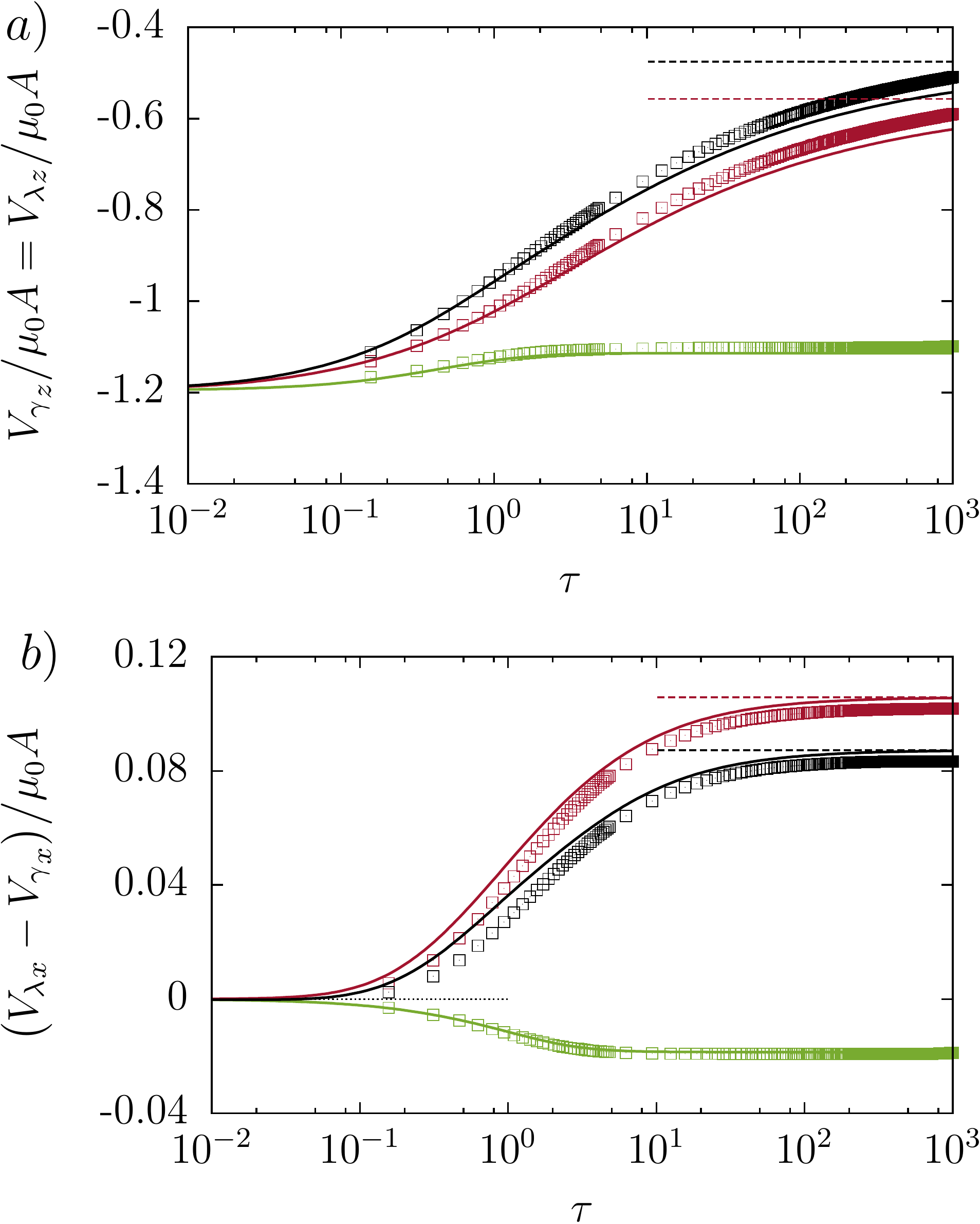}
     \caption{ (Color online) 
     The scaled particle velocities perpendicular to the membrane ($a$) and relative to each other ($b$) versus the scaled time for a constant force of amplitude $A$ acting downward on both particles near a membrane endowed with pure shear (green), pure bending (red) or both rigidities (black).
     Solid lines represent the analytical predictions given by Eq.~\eqref{velocityStepForce}. 
     Symbols refer to BIM simulations.
     Horizontal dotted and dashed lines stand for the bulk and vanishing frequency limits, respectively.
     From Daddi-Moussa-Ider and Gekle, \textit{J.~Chem. Phys.}, {\bf 145}, 014905 (2016), copyright 2016 American Institute of Physics.} 
     \label{attractionRepulsionGeschDown}
  \end{center}
\end{figure}

In order to elucidate the change of sign in the particle pair-mobility, 
it is judicious to consider the steady approach of two particles towards an elastic interface.
For hard walls, it is well known that hydrodynamic interactions are repulsive~\cite{dufresne00, squires00} leading to the dispersion of particles on the surface.
Near elastic membranes, the different signs of the shear- and bending-related contributions observed in the pair mobility (Fig.~\ref{mobiCorr}~$b$) point to a more complex scenario including the possibility of particle attraction.

The physical situation of a pair of particles being initially located at $z=z_0$ and suddenly set into motion towards an interface is described by a Heaviside step function force $\vect{F} (t) = \vect{A} \theta(t)$.
Using the general form of Eq.~\eqref{eqn:muGeneral}, the scaled particle velocity in the temporal domain is thus given by~\cite{daddi16c}
\begin{equation}
 \frac{V(\tau)}{\mu_0 A} = \left[ b + \int_{0}^{\infty} \frac{\varphi_1(u)}{\varphi_2(u)} \left( 1-e^{-\varphi_2(u) \tau} \right) \Intd u  \right] \theta (\tau) \, ,
 \label{velocityStepForce}
\end{equation}
where $\tau := t/T$ is a dimensionless time.
At larger times, the exponential in Eq.~\eqref{velocityStepForce} can be neglected compared to one and thus one recovers the steady-state behavior.

In corresponding BIM simulations, a constant force of small amplitude $A$ is applied on both particles towards the membrane.
For $\epsilon = 1/2$ and $\sigma = 1/4$, Fig.~\ref{attractionRepulsionGeschDown}~$a)$ shows the time dependence of the scaled vertical velocity, which at first increases and then approaches progressively its steady-state value. 
Fig.~\ref{attractionRepulsionGeschDown}~$b)$ shows the scaled relative velocity between the two particles.
Clearly, the motion is attractive for an idealized membrane with negligible bending resistance (such as a typical artificial capsule designed for drug delivery) which is indeed the opposite of the behavior near a membrane with pure-bending resistance (such as a fluid vesicle) or a hard wall.
For a membrane with pure shear, it can be shown that the threshold line where the shear-related contribution changes sign in the $xz$ component of the particle pair mobility function is given up to fifth order in $\sigma$ by~\cite{daddi16c}
\begin{equation}
 \ETH =  \sqrt{2}  \left( \sigma - \frac{4}{3} \, \sigma^3 + \frac{17}{27} \, \sigma^5  \right) + \bigO (\sigma^7) \, . \label{epsilon_XZ_line}
\end{equation}
For $\epsilon < \ETH$, the hydrodynamic interactions are repulsive whereas for $\epsilon > \ETH$ are attractive.

The following section addresses the Brownian motion of a particle diffusing near an elastic membranes. 
The aforementioned membrane-induced memory effects will result in a long-lived transient subdiffusion of the particle


\section{Brownian motion nearby membranes}\label{sec:brownianMotion}

\subsection{Generalized Langevin equation}

The frequency-dependent mobilities presented in the previous section can be used as input for the calculation of the particle diffusion tensor.
The dynamics of a Brownian particle in the presence of an elastic membrane is governed by a generalized Langevin equation~\cite{kubo66, ciccotti81}.
Restricting to one dimensional motion, we have
\begin{equation}
  m \, \frac{\Intd  V(t)}{\Intd  t} = -\int_{-\infty}^{t} \gamma(t-t')V (t') \, \Intd  t' + F (t) \, ,  \label{generalizedLangevinEqn}
\end{equation}
where $m$ is the particle mass, 
$V(t)$ is the particle translational velocity,
$\gamma(t)$ is the time dependent friction retardation function (expressed in kg/s$^{2}$) and $F(t)$ is a stochastic random force modeling the effect of the background noise caused by the fluid on the Brownian particle.
The random force is Gaussian distributed and satisfies the statistical properties 
\begin{align}
 \langle F (t) \rangle &= 0 \, , \\
  \langle F(t) F(t') \rangle &= 2 \gamma_0 \kBolt T \delta (t-t') \, ,
\end{align}
where brackets mean ensemble average, $\kBolt$ is the Boltzmann constant and $T$ is the absolute temperature of the system.
Here we assume that there are no other external forces acting on the particle.

In the particular case when $\gamma(t) = 2\gamma_0 \, \delta (t)$, with $\gamma_0 = \mu_0^{-1} = 6\pi\eta a$ being the bulk friction coefficient, Eq.~\eqref{generalizedLangevinEqn} is reduced to the classical non-retarded Langevin equation in which the random force are assumed to be a purely Gaussian process delta correlated in time.

The computation of the particle mean-square displacement (MSD) requires as an intermediate step the determination of the velocity autocorrelation function via the application of the fluctuation-dissipation theorem.

\subsection{Fluctuation-dissipation theorem}

Evaluating the Fourier transform of both members in Eq.~\eqref{generalizedLangevinEqn} as defined by Eq.~\eqref{fourierInTime}, we obtain
\begin{equation}
i m \omega V(\omega) = -\int_{-\infty}^{\infty} e^{-i\omega t} \, \Intd  t \int_{-\infty}^{t} \gamma(t-t') V(t') \, \Intd  t' + F(\omega) \, . \notag
\end{equation}
Using the change of variables $u=t-t'$, together with the shift property of Fourier transforms, the particle velocity is related to the fluctuating force via the frequency-dependent mobility as
\begin{equation}
	V(\omega) =  \mu(\omega) F(\omega) \, , \text{~with~}
	\mu(\omega) = \frac{1}{im\omega + \gamma[\omega]} \, , 
	\label{def-MU}
\end{equation}
wherein $\gamma[\omega]$ is the Fourier-Laplace (also called one-sided Fourier) transform of the retardation function, defined by
\begin{equation}
\gamma[\omega] = \int_{0}^{\infty} \gamma(t) e^{i \omega t} \, \Intd  t \, . \label{LaplaceFourierDef}
\end{equation}

In virtue of the fluctuation-dissipation theorem~\cite{kubo66}, the frictional forces and the random forces are not independent quantities, but they are rather related to each other via the correlation
\begin{equation}
\langle F(\omega) \overline{F (\omega')} \rangle = \phi_\mathrm{F}(\omega) \, \delta (\omega+\omega') \, , 
\end{equation}
where $\phi_\mathrm{F}(\omega)$ is the Fourier transform of the velocity autocorrelation function $\phi_\mathrm{F} (t)$, known also in the literature as the power spectrum of $F(\omega)$.
In term of the friction kernel, the power spectrum is given by
\begin{equation}
\phi_\mathrm{F} (\omega) = \kBolt T \left( \gamma[\omega] + \gamma[-\omega] \right) = 2 \kBolt T \operatorname{Re} \left( \gamma[\omega] \right) \, , 
\label{CF}
\end{equation}
noting that $\overline{\gamma[\omega]} = \gamma[-\omega]$, as can be inferred from Eq.~\eqref{LaplaceFourierDef}.
The power spectra of $V(\omega)$ and $F(\omega)$ are thus related to each other via the relation
\begin{equation}
\phi_\mathrm{V} (\omega) =  \frac{\phi_\mathrm{F} (\omega)}{|i m \omega + \gamma[\omega]|^2} \, .
\end{equation}

By making use of Eq.~\eqref{CF}, we obtain
\begin{equation}
\phi_\mathrm{V} (\omega) =  \frac{\kBolt T}{im\omega + \gamma[\omega]} + c.c. \, , 
\end{equation}
wherein $c.c.$ stands for complex conjugate.
Provided that $\gamma[\omega]$ is known, it is therefore possible to transform $\phi_\mathrm{V} (\omega)$ back to the time domain, leading directly to the velocity autocorrelation function
\begin{equation}
\phi_\mathrm{V}(t) := \langle V(0)V(t) \rangle = \frac{\kBolt T}{2\pi} \int_{-\infty}^{\infty} \left( \mu(\omega)+\overline{\mu(\omega)} \right) e^{i\omega t} \, \Intd  \omega \, ,
\label{CV}
\end{equation}
after making use of Eq.~\eqref{def-MU}.

From the general form of the mobility given by Eq.~\eqref{eqn:muGeneral}, it can clearly be seen  that the contribution from the second term in Eq.~\eqref{CV} vanishes for $t>0$ since the integrand has simple poles $\omega = -i\varphi_2(u) T$ and thus is analytic in the \emph{upper} half plane in which $\operatorname{Im} \omega >0$. 
As a result, the velocity autocorrelation function for $t>0$ reduces to~\cite{kubo85}
\begin{equation}
\phi_\mathrm{V} (t) = \frac{\kBolt T}{2\pi} \int_{-\infty}^{\infty} \mu (\omega) e^{i\omega t} \, \Intd  \omega \, .
\label{ACFDef}
\end{equation}

\subsection{Diffusion nearby cell membranes}

\begin{figure}
\begin{center}
 \includegraphics[width=8.5cm]{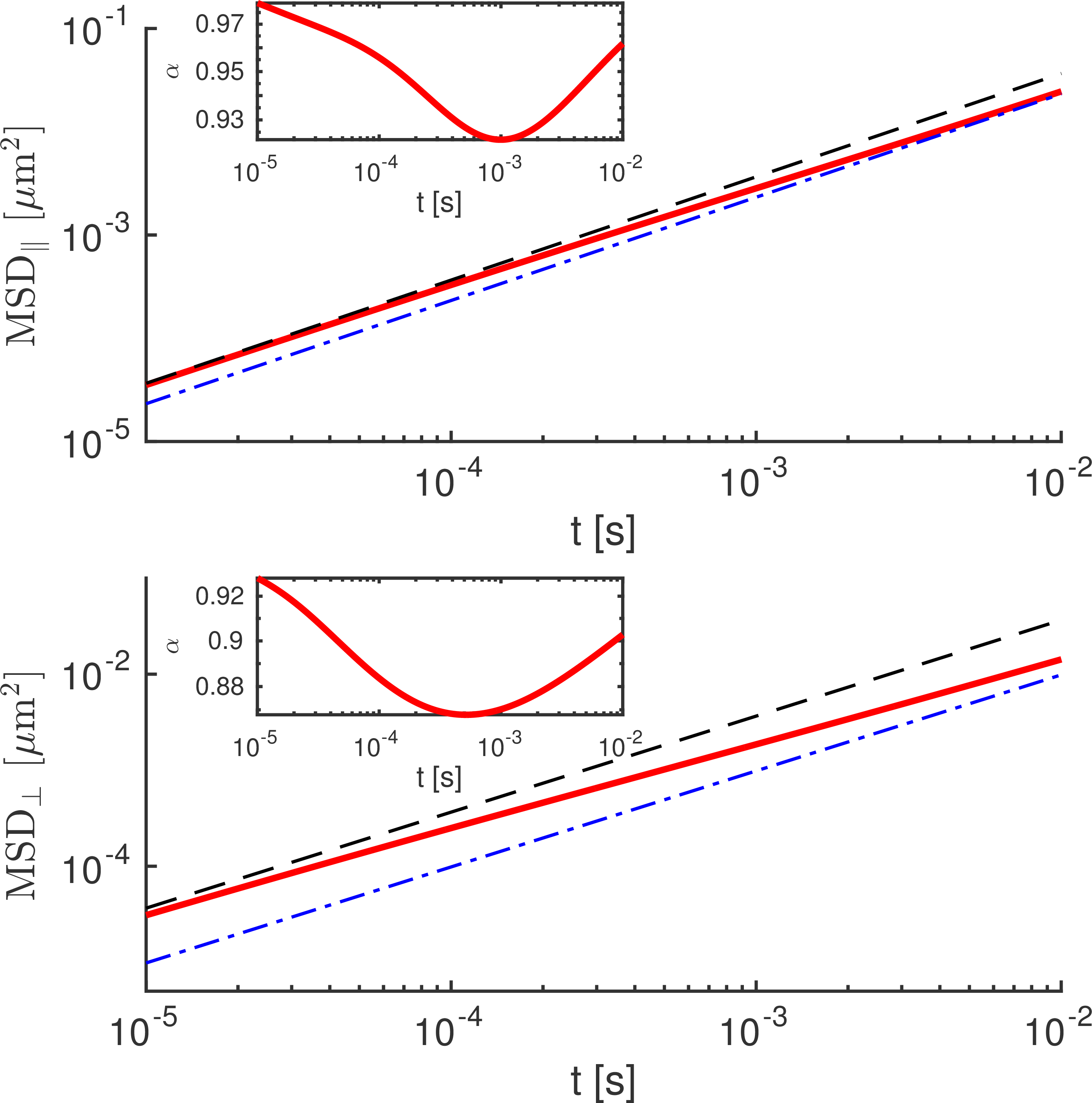}
\caption{Mean-square displacement (red line) of a Brownian particle of radius $a$=100nm, located at $z_0$=153nm above a RBC membrane in parallel (top) and perpendicular (bottom) direction as predicted by theory at $T=300$K. 
For short times $t\lesssim 50\mu$s the MSD follows bulk behavior (black dashed line) while for long times it follows hard-wall behavior (blue dash-dotted line). 
In between, a long-lived subdiffusive regime is evident extending up to 10ms and beyond. 
Insets show the local scaling exponent which goes down until 0.92 for parallel diffusion and 0.87 for perpendicular diffusion.
Adapted from Daddi-Moussa-Ider, Guckenberger and Gekle, \textit{Phys. Rev. E}, {\bf 93}, 012612 (2016), copyright 2016 American Physical Society.}
	\label{fig:MSD}
\end{center}
\end{figure}

Having computed the velocity autocorrelation function~$\phi_\mathrm{V}$, the MSD can thus be calculated from~\cite{kubo85}
\begin{equation}
\langle x(t)^2 \rangle = 2\int_{0}^{t} (t-s) \phi_\mathrm{V} (s) \, \Intd s \, .
\label{MSDDef}
\end{equation}
The particle long-time diffusion coefficient is computed as
\begin{equation}
	D_\infty := \lim_{t\to\infty}\frac{\langle x(t)^2 \rangle}{2t} = \int_{0}^{\infty} \phi_\mathrm{V} (s) \, \Intd s \, .
\label{DiffusionCoeffDef}
\end{equation}

An alternative way to quantify the slowing down of the particle and departure from standard diffusion is to investigate the time-dependent scaling exponent of the MSD, defined as the logarithmic derivative of the MSD such that
\begin{equation}
\alpha (t) := \frac{\Intd \ln \langle x (t) ^2 \rangle}{\Intd \ln t} \, .
 \label{scalingExponent} 
\end{equation}
If diffusion is normal (standard), then the scaling exponent is one.
Anomalous subdiffusion is characterized by a scaling exponent that is less than one and is often encountered in biological media with obstacles~\cite{saxton94} or binding sites~\cite{saxton96}.

A similar methodology can be adopted to investigate the diffusional dynamics of a pair of particles near a membrane, notably to calculate the collective and relative diffusion coefficients.
These have been considered in details in Ref.~\cite{daddi16c}.

As a system setup, a spherical particle with radius $a=100$nm is located at a distance $z_0=153$nm above a membrane and exhibits diffusive motion.
The elastic membrane is endowed with a shear modulus $\kappa_{\mathrm{S}}=5\times 10 ^{-6}$~N/m, a bending modulus $\kappa_{\mathrm{B}}=2\times 10 ^{-19}$~Nm, and a Skalak coefficient $C=100$ which are typical values for red blood cells~\cite{freund13}. 
The fluid properties correspond to blood plasma with dynamic viscosity $\eta=1.2$mPas.

Fig.~\ref{fig:MSD} shows the MSD for parallel as well as perpendicular diffusion as obtained from theory. 
For short times ($t<50~\mu$s) the MSD follows faithfully a linear behavior with the normal bulk diffusion coefficient $D_0 = \mu_0 \kBolt T$ given by Einstein's relation~\cite{einstein05} since the membrane does not have sufficient time to react on these short scales.
As the time increases, we observe a downward bending of the MSD which is a clear signature of an anomalous subdiffusive behavior. 
Indeed, as shown in the insets of Fig.~\ref{fig:MSD}, the local exponent may go as low as 0.92 in the parallel and as low as 0.87 in the perpendicular direction. 
The subdiffusive regime extends up to 10~ms in the parallel and even further in the perpendicular direction, which is long enough to be of possible physiological significance. 
Finally, for long times, the behavior turns back to standard diffusion with $\alpha\to 1$. 
Compared to the short-time behavior, however, the diffusion coefficient is now significantly lower and approaches the well-known behavior near a no-slip hard wall.
It turns out that diffusion for sufficiently long times to be universal and independent of the membrane properties.

\section{Conclusions}\label{sec:conclusions}

In this work, we have discussed the hydrodynamic interactions and Brownian motion near an elastic cell membrane possessing both shear and bending rigidities. 
The analytical methodology proceeds through the calculation of the Green's functions which are solutions of the linear Stokes equations due to a point-force singularity acting near the membrane. 
Thereupon, the hydrodynamic mobility functions which couple the particles' velocities to the forces applied on their surfaces have been determined.
For that purpose, a combination of multipole expansion and \Faxen~theorem has been employed to yield analytical expressions of the particle self- and pair-mobilities obtained directly from the Green's functions.
The corrections to the mobility functions have been expressed as a power series of the ratio between particle radius and distance from the membrane for the self-mobilities, and between particle radius and interparticle distance for the pair-mobilities.

It turned out that the particle mobilities near a planar membrane can appropriately be expressed as a linear superposition of the contributions stemming from shear and bending as obtained independently.
Moreover, the shear- and bending-related parts may have additive or suppressive contribution to the total mobility, depending on the membrane properties, the distance from the membrane in addition to the interparticle distance.
This interesting behavior has been elucidated by considering the startup motion of two particles suddenly set into motion towards a membrane.
The interaction near a membrane with pure bending resistance is always repulsive, \ie as in the case of a hard wall, while near a membrane with pure shear, the interaction may be attractive.
The theoretical predictions have been confirmed by full-resolved boundary integral simulations of truly finite-sized particles.

The calculation of the particle MSD characterizing the diffusion process reveals that elastic membranes induce a long-lived anomalous subdiffusion on nearby particles.
This behavior can significantly enhance residence time and binding rates nearby membranes and thus may increase the probability to trigger the uptake of particles via endocytosis.

\begin{acknowledgments}
The authors thank the Volkswagen Foundation for financial support and acknowledge the Gauss Center for Supercomputing e.V. for providing computing time on the GCS Supercomputer SuperMUC at Leibniz Supercomputing Center. 
We gratefully acknowledge support from the Elite Study Program Biological Physics and from the COST Action MP1305, supported by COST (European Cooperation in Science and Technology).
\end{acknowledgments}

%


\end{document}